# Crossover between Photochemical and Photothermal Oxidations of Atomically Thin Magnetic Semiconductor CrPS₄

Suhyeon Kim[1], Jinhwan Lee[2], Gangtae Jin[3], Moon-Ho Jo[3,4,5], Changgu Lee[2,6], and Sunmin Ryu[1,3]*

[1]Department of Chemistry, Pohang University of Science and Technology (POSTECH), Pohang 37673, Korea

[2]School of Mechanical Engineering, Sungkyunkwan University, Suwon 16419, Korea

[3]Division of Advanced Materials Science, Pohang University of Science and Technology (POSTECH), Pohang 37673, Korea

[4]Center for Artificial Low Dimensional Electronic Systems, Institute for Basic Science (IBS), Pohang 37673, Korea

[5]Department of Materials Science and Engineering, Pohang University of Science and Technology (POSTECH), Pohang 37673, Korea

[6]SKKU Advanced Institute of Nanotechnology (SAINT), Sungkyunkwan University, Suwon 16419, Korea

## ABSTRACT

Many two-dimensional (2D) semiconductors represented by transition metal dichalcogenides have tunable optical bandgaps in the visible or near IR-range standing as a promising candidate for optoelectronic devices. Despite this potential, however, their photoreactions are not well understood or controversial in the mechanistic details. In this work, we report a unique thickness-dependent photoreaction sensitivity and a switchover between two competing reaction mechanisms in atomically thin chromium thiophosphate (CrPS₄), a 2D antiferromagnetic semiconductor. CrPS₄ showed a threshold power density two orders of magnitude smaller than that for MoS₂ obeying a photothermal reaction route. In addition, reaction cross section quantified with Raman spectroscopy revealed distinctive power dependences in the low and high power regimes. Based on optical in-situ thermometric measurements and control experiments against O₂, water, and photon energy, we proposed a photochemical oxidation mechanism involving singlet O₂ in the low power regime with a photothermal route for the other. We also demonstrated a highly effective encapsulation with Al₂O₃ as a protection against the destructive photoinduced and ambient oxidations.

## 1. Introduction

Various materials properties of two-dimensional crystals (2DXs) are affected by structural defects that are native or induced by external perturbation.[1-5] Because of such pivotal role of defects, chemical stability of 2DXs with predesigned properties is of prime concern not only to their fundamental research but also to device applications in switching electronics, energy, sensors, optoelectronics, and catalysis.[1, 3, 5] Since the isolation of graphene in 2004,[6] tens of 2DXs have been investigated for various scientific studies and application purposes, which resulted in tens of thousands of research papers. Some including hexagonal-BN, graphene and $MoS_2$ are stable in ambient conditions and can be isolated by physical exfoliation from their bulk crystals because of weak van der Waals (vdW) attraction between their 2D building blocks. Unfortunately, however, many 2DXs such as transition metal dichalcogenides (TMDs),[7] $CrI_3$,[8] and phosphorene[9] lack stability even in the ambient air and require careful handling and encapsulation of exposed surfaces to minimize chemical degradation. Compared to their bulk forms, some 2DXs are more vulnerable to chemical attacks because of thickness-dependent bandgaps,[9] structural deformation,[10] and substrate-mediated charge imbalance.[11] Because of the widely-varying susceptibility of 2DXs towards ambient chemical reactions, generalized understanding requires more systematic and extensive research efforts. In this regard, a newly emerging 2DX family originating form layered ternary transition metal chalcogenides (TTMCs)[12] in the form of $MAX_n$ (M = transition metal such as Mn, Fe, Ni, Cr, Co; A = P, Si, Ge; n = 3, 4) will serve as an excellent testbed because of their large number of family members that allow systematic and comparative investigation on how geometric and electronic structures affect their chemical properties. In addition, atomically-thin TTMCs[13, 14] bear not only scientific but also technological importance as candidates for 2D magnetic semiconductors[14, 15] since many bulk TTMCs including chromium thiophosphate ($CrPS_4$)[13, 16] studied in this work exhibit ferromagnetism or antiferromagnetism.

In addition to ambient stability, 2DXs are required to be robust under light to be considered for various functions, since photons are a ubiquitous perturbation even present in spectroscopic probes or room light, and can be detrimental to otherwise stable materials because of their high quantum of energy. One photon of green color, for instance, contains ~100 times of the thermal-energy quantum at room temperature and thus may drive otherwise improbable reactions with sizable activation energies. Because of the extremely high surface-to-volume ratio, photoreactions of 2DXs occur essentially on their surfaces and thus can be understood using established surface photochemistry.[17] First of all, intense laser radiation may increase local temperature sufficiently high and induce various thermal reactions including oxidation, which is referred to as a photothermal reaction (Figure 1a). Photooxidation of $MoS_2$ belongs to this mechanism,[18] which can also be operative to many 2DXs. In a direct photodissociation (Figure 1b), photons impinging on a 2DX excite the system onto a repulsive potential energy surface along an associated reaction coordinate, which is followed by scission of its lattice bonds. While photoinduced halogenation of graphene is initiated by direct photodissociation of halogen molecules,[19] no 2DX has been reported to follow this mechanism to the best of our knowledge. Photoexcited electrons (or holes) of the 2DX are transferred to another chemical entity that undergoes a bond-breaking reaction in a hot electron-mediated mechanism (Figure 1c). Photodegradation of graphene by benzoyl peroxide[20] and photooxidation of 2D black phosphorus[9] in ambient air are initiated by generation and subsequent transfer of hot electrons. In a sensitized photodissociation (Figure 1d), energy primarily absorbed in the 2DX may be transferred to an adsorbed molecule through sensitization with or without spin-flip.[21, 22] Subsequently the excited molecule may participate in a reaction alone or with the 2DX. While nano-sheets of black phosphorus sensitize dissolved oxygen molecules efficiently,[23] ensuing reactions of the singlet $O_2$ with the nanosheets were not studied. Despite the vast amount of research efforts, however, mechanistic studies of photoreactions of 2DXs have been very scarce as summarized above and far from reaching a mature stage. Moreover, the prior studies reported on a single mechanism that is operative within a narrow range of photo-perturbation, and thus failed to reveal how multiple reaction routes compete under a low or high flux of photons.

In this work, we show that few-layer $CrPS_4$, a semiconducting antiferromagnet, exhibits a photoreaction threshold two orders of magnitude lower in power density than $MoS_2$, and undergoes dual competing photooxidations each driven by photosensitization of oxygen molecules and photothermal heating. To quantify the degree of reaction and local temperature, we exploited Raman spectroscopy in conjunction with atomic force microscopy (AFM). In a low-power regime where lattice temperature was maintained below a threshold for thermal oxidation, the reaction rate was virtually independent of power density indicating that the reaction is a one-photon process. With increasing power density in a high-power regime, the lattice temperature rose, and the reaction rate surged sharply. As a reference that undergoes photothermal oxidation, few-layer $MoS_2$ was comparatively investigated. This work constituting the first report on a crossover between two competing reaction mechanisms will benefit future studies and applications of 2DXs.

## 2. Methods

**Preparation and characterization of samples.** Bulk CrPS$_4$ crystals were grown by the chemical vapor transport method as recently reported.[13] Bulk MoS$_2$ crystals, mainly of 2H polytype, were used as obtained from a commercial source (SPI). Single and few-layer CrPS$_4$ and MoS$_2$ samples were prepared by mechanical exfoliation of the bulk crystals.[24] Unless otherwise noted, silicon wafers with 285 nm thick SiO$_2$ were used as substrates. To avoid optical artifacts induced by multiple reflections and interference from substrates, we prepared some samples on amorphous quartz slides (SPI, SuperSmooth). To minimize degradation in the ambient air, we stored prepared samples in a dark vacuum desiccator maintained below 25 Torr. Thickness and quality of prepared samples were characterized by Raman spectroscopy, optical reflectance and atomic force microscopy (AFM).[13] Height and phase AFM images were recorded with a commercial unit (Park Systems, XE-70) in the non-contact mode using silicon tips with a tip radius of 8 nm (MicroMasch, NSC-15).

**Raman measurements.** Raman spectra were obtained with a home-built micro-Raman setup described in detail elsewhere.[25] An excitation laser beam, 514.3 nm in wavelength unless otherwise noted, was focused on samples using an objective lens (40X, numerical aperture = 0.60). Back-scattered Raman signals collected by the same objective lens were guided into a Czerny-Turner spectrometer (Princeton Instruments, SP2300) that was connected to a CCD camera (Princeton Instruments, PyLon). In experiments for dependence on photon energy, two additional laser lines of 457.0 and 632.8 nm were employed. While the spectral resolution defined by FWHM of a Rayleigh peak was 3.0 cm$^{-1}$, the spectral accuracy was better than 0.5 cm$^{-1}$. To avoid orientation-dependent variation of Raman signals, we maintained the polarization of the excitation beam parallel to the b axis (Figure 2a) that could be readily determined under an optical microscope.[13]

**Optical reflectance measurements.** To determine thicknesses of few-layer CrPS$_4$ and MoS$_2$ samples, we obtained their optical contrasts using an optical microscope (Nikon, LV100; 100X objective lens) equipped with a CMOS camera. The red (blue)-channel optical contrast, defined as the fractional change in reflection of red (blue) color with respect to bare SiO$_2$/Si (quartz) substrates, gave the best sensitivity and reliability in resolving thickness.[13]

**Photoirradiation and in-situ monitoring.** Raman excitation beam was used as a light source for controlled photoirradiation and its effects were characterized by Raman spectra obtained simultaneously. To determine photon fluence for the three wavelengths, we measured the average power of the irradiation beam at a sample plane with a photodiode-type power meter. The effective beam diameter at the focus was estimated using a knife-edge method (Supporting Information Section B). For environment-controlled experiments, samples were placed in a gas-tight optical cell with independent controls over the mole fraction of oxygen in Ar, $\chi_{ox}$, and relative humidity (RH), respectively. Flow rates for $\chi_{ox}$ = 0.20 were 50 and 200 mL/min for $O_2$ and Ar gas, respectively, and varied accordingly for other $\chi_{ox}$ values. Water vapor was introduced to the cell by sparging the argon gas through deionized water.

**Encapsulation by ALD method.** $Al_2O_3$ thin films were deposited on 4L $CrPS_4$/$SiO_2$/Si using atomic layer deposition (ALD) as an encapsulation layer. Trimethylaluminum (TMA) and $H_2O$ vapor were fed into a reactor respectively as a precursor and an oxidant for deposition of $Al_2O_3$.[26] Each cycle of ALD consisted of four steps: feeding TMA for 0.3 s, purging Ar for 15 s, feeding $H_2O$ for 0.2 s and purging Ar for 15 s. When conducted at 100 °C, each cycle led to growth of 0.084 nm. The base pressure of the reactor was ~1 Torr with the presence of $N_2$ carrier gas (99.9999%) at a flow rate of 50 mL/min. Prior to deposition, samples were rapidly annealed for 120 s at 100 °C ($N_2$ flow rate at 500 mL/min) to minimize possible surface contaminants.

## 3. Results and Discussion

**Raman spectra of single and few-layer CrPS₄.** As described in the Methods section, single and few-layer $CrPS_4$ samples were prepared by mechanical exfoliation of bulk crystals grown by chemical vapor transport method.[13] Bulk $CrPS_4$ belongs to a space group of $C_3^2$ in the monoclinic family with a = 10.871 Å, b = 7.254 Å, c = 6.140 Å, β = 91.88°.[27] As shown in Fig. 2a, each Cr atom is coordinated by six S atoms arranged in a distorted octahedron and three S-octahedrons are connected by a single P atom that is located at the center of an S-tetrahedron. While sulfur atoms terminate each surface of a given single layer, they are not located in the same plane unlike chalcogens in 2H-type transition metal dichalcogenides (TMDs).[28, 29] Exfoliated samples typically contained terraces spanning several microns (see the optical micrograph in Fig. 2b), and their surfaces were very smooth without noticeable structural irregularities as shown in the AFM image in Fig. 2c. The smallest step height of 6.8 Å corresponded to the lattice constant c (or precisely $C \cdot \sin\beta$) which indicated that exfoliation occurred in a unit of a single layer (1L). In order to determine the number of layers and quality for each sample, Raman spectroscopy was employed according to the previous work.[13] As shown in Fig. 2d, 1L ~ 5L $CrPS_4$ samples placed in an Ar-filled optical cell exhibited several characteristic peaks in the range of 100 ~ 450 $cm^{-1}$. Raman peaks A, B, C, F, I and L for 4L $CrPS_4$ appeared at 116, 155, 170, 257, 306 and 409 $cm^{-1}$, respectively (labels of the peaks were made according to the previous work[13]). Unlike the previous work, we were able to obtain Raman spectra from 1L samples but only after prompt and careful encapsulation with $Al_2O_3$ layers (see Methods). As will be discussed below, this indicated that thinner $CrPS_4$ is even more susceptible to chemical degradation that occurs in the ambient air.

As shown in Fig. 2(d~f), the Raman spectrum of 1L was distinct from those from thicker layers for its large linewidths. In addition, peak frequencies of some peaks were significantly different from their bulk counterparts. The frequency of Raman peak B (L) decreased (increased) with decreasing thickness (Fig. 2e), which agreed with the previous results.[13] Raman peaks A and B, in particular, downshifted ~6% for 1L from those of the bulk (Fig. S1). When 2L $CrPS_4$ is considered as a vdW-coupled oscillator consisting of two 1L's, its out-of-phase normal mode has a higher resonance frequency than that of 1L because of the increased restoring force.[30] Thus, the opposite trend observed for L peak can be attributed to the surface relaxation,[30, 31] which leads to a mode hardening for outermost layers with respect to inner ones because of fewer adjacent layers. The surface effect will be more outstanding in Raman spectra of thinner layers as seen in Fig. 2(d~f). Resolving the contributions of the two opposing effects will require normal mode analysis and further theoretical study. However, it is to be noted that the frequency difference between peaks L and B ( $\omega_L$ - $\omega_B$ ) can serve as a reliable indicator for the thickness of few-layer samples (Fig. 2f) as for $MoS_2$ and other 2D TMD crystals.[30] The larger linewidths for thinner layers (Fig. S1) originate from the confinement effect of the associated phonons, which relaxes the wave vector selection rule at the Brillouin zone center.[31] One also needs to consider inhomogeneous broadenings induced by the chemical degradation and structural deformation affected by the substrates,[32] both of which are more likely in thinner samples. Despite the peak broadening, however, Raman intensities ( $I$ ) were proportional to the thickness as shown in Fig. 3a for peak F ($I_F$) (see Fig. S1 for other peaks), which was used below in quantifying the amount of intact $CrPS_4$ during chemical reactions.

**Distinctively low photoreaction threshold of 2D $CrPS_4$.** As suggested by the ambient instability of 1L, we found that $CrPS_4$ is much more susceptible to photoinduced reactions than TMDs such as $MoS_2$, $MoSe_2$, $WS_2$, and $WSe_2$. To judge a threshold power density, we investigated topographic details of samples that were photoirradiated in the ambient air with 514 nm laser beam by varying average power. Power density was calculated using the beam diameter (Fig. S2 & Table S1) determined by the knife edge method assuming 2D Gaussian intensity profile at the focus (see SI Section B). For the 514 nm excitation, $3.4 \times 10^5$ W/cm² in power density and $8.9 \times 10^{26}$ /cm²·s in photon flux corresponded to 1 mW in average power, which will be used below for simplicity. The AFM height images in Fig. 3a showed that the irradiated areas of 4L $CrPS_4$ inflated by 0.2 nm for 5 µW and the morphological change became more evident for higher powers. In contrast, the surface of 4L $MoS_2$ samples did not exhibit any noticeable change at powers lower than 0.3 mW but showed protrusions of 1 ~ 2 nm in height (Fig. 3b) at higher powers. The photogenerated structures on $MoS_2$ surfaces in Fig. 3b are Mo oxides formed through photothermal oxidation as will be corroborated below. Despite the apparent morphological difference in their photo-products, the change found in $CrPS_4$ is also oxidation as will be supported and explained later. However, the fact that the photooxidation threshold for $CrPS_4$ is 20 times lower than that for $MoS_2$ implies that the former may follow a reaction mechanism different than the photothermal channel.

**Crossover between photochemical and photothermal mechanisms.** In order to unveil the photoreaction mechanisms of $CrPS_4$, we performed quantitative analyses of reaction kinetics using its Raman spectra obtained in various conditions. Figure 4c presents Raman spectra consecutively obtained from one spot of a 4L $CrPS_4$ sample, where the Raman excitation beam itself induced the photoreaction. For higher Raman signals, 4L samples were selected as a representative system while similar results were obtained for thinner samples as will be described later. With increasing photon fluence ( $\Phi$ ), all the peaks including the peak F decreased in intensity indicating oxidation of $CrPS_4$. With increasing $\Phi$, $I_F$ normalized by that of intact 1L decreased almost linearly at the beginning and the rate of decrease slowed down significantly for $I_F <$ 1 (Fig. 3d). Since $I_F$ is proportional to the amount of intact $CrPS_4$ in a focal volume (Fig. 2d), the negative value of the tangential slope to the data in Fig. 3d essentially represent the rate of photooxidation at a given photon flux ( $\varphi$ ).

For quantitative understanding, a simple kinetic model was set up for a photochemical reaction of a multilayered 2DX sample (thickness of $D_0$ ) supported on a transparent substrate. Monochromatic light of $\varphi$ impinges on the sample surface with zero incidence angle. It is assumed that the reaction is surface-localized only consuming the top-most layer and that products of the reaction do not interfere with the reaction. Assuming that the reaction obeys a modified first-order kinetics, then the effective thickness ( $D$ ) of unreacted 2DX satisfies the following rate equation until the reaction starts to expose the bare substrate:

$$\frac{dD}{dt} = -\sigma \cdot \varphi \cdot D_0^{1L}$$ , where $\sigma$ and $D_0^{1L}$ are a reaction cross-section and the thickness of intact 1L,

respectively.[33] Then, the product $\sigma \cdot \varphi$ represents an effective rate constant ($k$). It is to be noted that the rate of a photothermal reaction will be highly nonlinear in $\varphi$ since the Arrhenius-type rate constant depends on temperature exponentially. The photochemical rate is linearly dependent on $\sigma$ that is, in general, a function of photon energy, surface structure, and composition. The rate is also proportional to the constant $D_0^{1L}$, not $D$ in an original first-order kinetics since the reaction occurs only at the surface layer. When transformed into an integrated form, the rate equation becomes: $\dfrac{D_0 - D}{D_0^{1L}} = \sigma \cdot \varphi \cdot t = \sigma \cdot \Phi$ In the right ordinate of Fig. 3d, the normalized extent of reaction ($\xi = \dfrac{D_0 - D}{D_0^{1L}}$) is given as a function of $\Phi$. Since the slope of its tangent representing $\sigma$ decreased as the reaction proceeded, we defined an initial reaction cross-section ($\sigma_i$) as the average slope over the data range of $0 < \xi < 1$ as marked by the red dotted line in Fig. 3d. For a 4L sample, $\sigma_i$ then refers to the oxidation of its top 1L or its equivalent.

In Fig. 4a, we monitored the photoreaction of 4L CrPS$_4$ as a function of $\Phi$ by varying the average power of 514 nm beam in the range of 0.025 ~ 10 mW. As indicated by the topographic changes in Fig. 3a, 4L CrPS$_4$ underwent photooxidation at average powers much lower than 1 mW, and its rate surged at 10 mW. As can be seen in Fig. 4c, $\sigma_i$ exhibited two distinctive power-dependences that were divided around 3 mW. In the low power regime (< 3 mW), $\sigma_i$ showed a very weak dependence on power. The reaction was significant even at the minimum power of 0.025 mW required for the Raman measurements and $\sigma_i$ increased only by ~110% when average power was 40 times increased to 1 mW. In contrast, a further increase to 8 mW led to 5-times increase in $\sigma_i$. These results suggest that two different mechanisms are operative in the low and high-power regimes, respectively. Below we will show that CrPS$_4$ undergoes photochemical and photothermal oxidations with the latter favored at elevated temperatures and that the crossover between the two occurs near 3 mW or power density of $1.0 \times 10^6$ W/cm$^2$.

**Confirmation of photothermal mechanism by Raman thermometry.** As a reference oxidation system obeying the photothermal mechanism, 4L MoS$_2$ was investigated for its kinetic behaviors (Fig. 4b). When the average power was below 6 mW, its $E_{2g}^1$-Raman intensity representing the effective thickness of intact MoS$_2$ did not change even when $\Phi$ reached 4 times of what was required to consume 4L CrPS$_4$ mostly in the low-power regime (see Fig. S3 for their Raman spectra). Above 6 mW, noticeable photooxidation was observed and $\sigma_i$ rapidly increased with increasing average power as shown in Fig. 4c. Compared to CrPS$_4$, $\sigma_i$ of MoS$_2$ was 3 orders of magnitude smaller near the threshold power and the difference decreased at the high power regime. It is to be noted, however, that the first-order kinetics model giving $\sigma_i$ is not relevant to MoS$_2$ but only for the sake of comparing reactivity since the system follows the photothermal route.

In order to unravel the highly nonlinear power-dependence of $\sigma_i$, photoinduced temperature rise was determined using Raman thermometry,[34] where one measures anti-Stokes/Stokes intensity ratio that is governed by the Boltzmann distribution of vibrational states[35] (SI Section D). The sample temperature ($T_{opt}$) was optically averaged over a focal spot and calibrated with respect to a room temperature of 25 °C by extrapolating to a zero power (Fig. S4). For 4L MoS$_2$, $T_{opt}$ determined with $E_{2g}^1$ mode remained below 70 °C at 1 mW and still below 150 °C at 4 mW (Fig. 4d), which is consistent with the negligible reaction in Fig. 4d considering that the threshold temperature for thermal oxidation is above 300 °C.[36-38] When the average power was raised to 11 mW for which $\sigma_i$ reached its maximum (Fig. 4c), $T_{opt}$ reached 320 °C where thermal oxidation readily occurs. Since $T_{opt}$ represents the temperature averaged over the entire focal spot, the peak temperature at its center can be significantly higher. Indeed a recent numerical simulation predicted that the peak temperature could be ~100 °C higher than the average value of 150 °C for a 532 nm-irradiated laser spot SI Section D.[18] Thus our in-situ temperature measurements confirmed that the photooxidation of MoS$_2$ is thermally-driven. The difference in the threshold power with respect to the previous work (3.7 mW)[18] can be attributed to differences in focal spot sizes, structural qualities of samples, substrates, excitation energy and thickness of samples.

The photoreaction of CrPS$_4$ in the high-power regime was also characterized by monitoring its temperature determined with its Raman F mode (Fig. 4d; Fig. S4 for Raman spectra). While the temperature rise was negligible below 1 mW (low-power regime), $T_{opt}$ rose to ~100 and 160 °C at 6 and 14 mW, respectively. It can be seen that the crossover between the two reaction routes occurred at $T_{opt}$ = 50 ~ 70 °C. To determine a threshold temperature for thermal oxidation, we oxidized 4L CrPS$_4$ samples in a tube furnace at various temperatures. After each step of oxidation, the effective thickness of intact CrPS$_4$ was determined by the Raman signal $I_F$. As indicated by the sharp decrease of $I_F$ in Fig. 4e, thermal oxidation occurred at or above 200 °C and consumed 4L within 1 hr at 250 °C. The optical contrast obtained from their optical micrographs (Fig. S5) also changed drastically near the threshold temperature (Fig. 4e). Noting that the peak temperature at the focus can be much higher than $T_{opt}$, we conclude that the crossover temperature approximately corresponds to the threshold for photoinduced thermal oxidation and that the reaction in the high-power regime is mainly driven through the photothermal channel. It is also to be noted that the measured temperature for a given power above 1 mW was $40 \sim 100$ ºC lower for $CrPS_4$ than $MoS_2$ on average. The difference can be partly because $CrPS_4$ has 4.5 times smaller absorption than $MoS_2$ as shown in their absorptance data (Fig. S6). The photoinduced rise in temperature will also depend on specific heat and thermal dissipation.[18] In order to estimate thermodynamic energy barriers for the photothermal reactions, the effective rate constants ( $k$ ) for the high-power regime were analyzed in the Arrhenius-type plots in Fig. 4f. The activation energies determined from the slopes were 17.5 and 109 kJ/mol respectively for $CrPS_4$ and $MoS_2$, confirming the higher chemical fragility of the former. We also note that the value of $MoS_2$ is in a reasonably good agreement with $77 \sim 153$ kJ/mol for dissociative adsorption of $O_2$ from a recent theoretical prediction,[39] where the lower value is associated with the presence of S vacancy.

**Nature of photochemical reaction of $CrPS_4$.** In order to identify the low-power photoreaction of $CrPS_4$, a few environmental factors were systematically tested below. First of all, the reaction required oxygen molecules. As the mole fraction of $O_2$ ( $\chi_{ox}$ ) in the optical gas cell was increased, the reaction was accelerated (Fig. 5a). As shown in Fig. 5b, $\sigma_i$ increased rapidly with increasing $\chi_{ox}$ up to 0.4 and the increase became modest showing a saturation behavior at higher $\chi_{ox}$. Secondly, a possible role of water vapor was studied by varying the relative humidity (RH) in the cell with $\chi_{ox}$ fixed at 0.20. As can be seen in Fig. 5c and 5d, $\sigma_i$ did not show any statistically meaningful deviation in the RH range of $12 \sim 84\%$. The insensitivity of $\sigma_i$ to water vapor suggests that the reaction is not mediated by charge transfer (CT) photooxidation that was proposed for the ambient oxidation of black phosphorus.[9] The model suggested that photogenerated superoxides ($O_2^-$) are the oxidants and also implicitly assumed that black phosphorus surface and $O_2$ are hydrated. The photoexcited electrons in black phosphorus are transferred to the electron affinity level ($E_{ox}$ in Fig. 1c) of hydrated $O_2$ that is located 3.1 eV below the vacuum level.[40] Noting that the electron affinity of an isolated $O_2$ molecule is 0.45 eV,[41] significant stabilization of the ionic species by hydration is essential for the CT process, which is consistent with the fact that the CT photooxidation required water vapor in addition to $O_2$.[9] However, we cannot exclude the possibility that water plays a certain role because of the limited range over which RH could be controlled. Thirdly, the photoreaction occurred in a wide range of excitation energy. As shown in Fig. S7, $\sigma_i$'s determined at 1.96, 2.41 and 2.71 eV were found to be directly proportional to the degree of absorption.

As an alternative to the photothermal and CT-mediated mechanisms, we propose that the reaction in the low-power regime is photochemical oxidation that is initiated by photosensitization of $O_2$ by $CrPS_4$ and completed when photogenerated singlet $O_2$ attacks $CrPS_4$ as follows:

$$CrPS_4 + h\nu \rightarrow CrPS_4^* \qquad (1)$$

$$CrPS_4^* + O_2\left(^3\Sigma_g^-\right) \rightarrow CrPS_4 + O_2^*\left(^1\Delta_g\right) \qquad (2)$$

$$O_2^* + CrPS_4 \rightarrow \left(CrPS_4 \cdot O_2\right) \rightarrow\rightarrow CrPO_4 + SO_2 \qquad (3)$$

The first step of the reaction represented as Equation 1 above is the spin-allowed optical absorption mainly localized at $Cr^{3+}$ ions in the octahedral ligand fields.[13, 42] J. Lee et al. reported that few-layered $CrPS_4$ has two broad absorption bands centered at 1.7 and 2.8 eV in the visible region, which correspond to $^4A_{2g} \rightarrow {}^4T_{2g}$ and $^4A_{2g} \rightarrow {}^4T_{1g}$, respectively.[13] Part of the absorbed energy is emitted as photoluminescence centered at 1.3 eV.[13]

In the second step of sensitization (Equation 2), photoexcited $CrPS_4$ is deactivated by colliding $O_2$ in its ground triplet state, which is subsequently excited to its singlet state.[43] A similar sensitized generation of singlet $O_2$ by Cr(III) complexes has recently been reported.[44] In the ligand field theory, the low-energy photophysics of Cr(III) complexes and solids are mainly governed by the excited quartet ($^4T_{2g}$) and doublet ($^2E_g$) states, the energy ordering of which is determined by the strength of the ligand fields.[45] While detailed energetics has yet to be revealed for $CrPS_4$, it is known that their energy difference is not significant that intersystem crossing between the two states is efficient for a wide range of ligand field strength.[46] This suggests that the PL emission observed in the previous report[13] may contain phosphorescence from the doublet state in addition to the fluorescence from the quartet state. Moreover, the energy transfer in the second step is energetically favorable since the PL energy of $CrPS_4$ (1.3 eV) is sufficiently larger than the energy gap (0.977 eV) between the triplet and the singlet states of $O_2$.[47]

The last step in Equation 3 represents the oxidation of $CrPS_4$ surface by singlet $O_2$. Compared to its ground-state triplet form, singlet $O_2$ is much more reactive towards various chemical species and induces reactions such as Diels-Alder cycloaddition; formation of sulfoxides, hydroperoxides, endoperoxides; various cytotoxic effects; and activation of gene expression.[48, 49] It also transforms various organic sulfides ($R_2$-S) via intermediates of persulfoxides ($R_2$-S-O-O).[50, 51] Each S atom of 1L $CrPS_4$ is bonded to one P and $1 \sim 2$ Cr atoms and all S atoms are exposed on either of its two surfaces. It is likely that persulfoxides also mediate the oxidation of $CrPS_4$. Since singlet $O_2$ is exceptionally long-lived owing to the spin-forbidden radiative relaxation, for example ~1 min in vacuum[47] and 50 ~ 90 ms in the air,[21] its encounter with $CrPS_4$ is highly plausible. While the detailed reaction mechanism of Equation 3 is not known currently, its solid product is likely to be $CrPO_4$ as in the thermal oxidation reaction.[16] Theoretical study on the reaction dynamics going beyond the scope of the current work will be needed for complete understanding of the reactions.

**High chemical fragility of 1L $CrPS_4$ and its protection by encapsulation.** We could not detect any significant Raman signal from freshly prepared 1L samples in an inert Ar gas, unlike multilayers. This observation implied that a minimum photon exposure required for one Raman spectrum of multilayers was detrimental enough to destroy 1L completely. The thickness-dependence of $\sigma_i$ in Fig. S8 showed that the rate of photochemical oxidation is at least twice higher for 1L than multilayers but similar among multilayers of 2L ~ 5L. The distinctively high reactivity of 1L may originate from thickness-dependent structural disorder enforced by strong conformal adhesion with underlying substrates.[32] In general, lattice sites with high local strain or defects will present smaller activation energy for a given chemical reaction. Similar observations were made for thermal oxidation,[10] hydrogenation[52] and photochemical oxidation[20] of graphene. While differences in electronic structures and dynamics of excited charge carriers may contribute, information currently available[13] is not sufficient for clear understanding.

Finally, we tested encapsulation with optically transparent $Al_2O_3$ layers as a photo-protection method. As shown in Fig. 5e and 5f, 5 nm or thicker $Al_2O_3$ grown on top of 4L $CrPS_4$ by atomic layer deposition (ALD)[26] exhibited oxygen-blocking effect[53] that is equivalent to that of Ar-purging. A similar degree of photostability was also confirmed for encapsulated 1L samples in the ambient air (Fig. S9). In addition, the encapsulated 1L samples were found to be stable after prolonged storage in the dark vacuum desiccator for 5 months. Noting the fact that unprotected 1L samples underwent complete photodegradation immediately even in a high purity Ar gas as mentioned regarding Fig. 2d, $Al_2O_3$ passivation layers are highly effective in blocking diffusion of $O_2$ molecules. We also note that freshly prepared 1L samples were degraded by the ambient oxidation on a time scale of a few weeks when judged from the encapsulation tests of samples that were pre-aged for various periods in the dark vacuum desiccator (Fig. S10).

**4. Conclusions**

In this work, we studied photoreaction mechanisms of atomically thin CrPS$_4$, a layered antiferromagnetic semiconductor and showed that two competing reaction channels switch over depending on radiant power density. Compared to thin MoS$_2$ serving as a reference obeying a photothermal oxidation mechanism, CrPS$_4$ exhibited two orders of magnitude smaller threshold power density. Reaction kinetics quantified by Raman spectroscopy showed that the reaction cross section was nearly power-independent in a low power regime but rapidly increased for a high power regime, which indicated the presence of multiple reaction routes. Optical in-situ thermometry and independent thermal reaction tests revealed that photochemical and photothermal mechanisms are operative in the respective power regimes. Through systematic control experiments against oxygen, water vapor and excitation photon energy, a detailed mechanism for the photochemical route involving generation of singlet oxygen was proposed. In addition, we showed that 1L CrPS$_4$ is even more fragile than multilayers towards photochemical and ambient oxidations. As a practical remedy, we demonstrated passivation by Al$_2$O$_3$ layers, which allowed the first-time detection of Raman spectrum from 1L samples.

## ASSOCIATED CONTENT

**Supporting Information.** Raman peak intensities, frequencies and linewidths of 1L ~ 5L and bulk CrPS$_4$, Characterization of beam intensity profile using knife-edge method, Raman spectra of 4L MoS$_2$ and 4L CrPS$_4$ obtained in-situ during photoreactions, Raman thermometry for 4L MoS$_2$ and 4L CrPS$_4$, Optical contrast change in 4L CrPS$_4$ during thermal oxidation, Differential reflectance spectra of 4L CrPS$_4$ in comparison with 4L MoS$_2$, Photon energy dependence of photooxidation for 4L CrPS$_4$, Thickness dependence of photooxidation for CrPS$_4$, Effects of extended storage after and pre-aging prior to encapsulation, and supporting references. This material is available free of charge via the Internet at http://pubs.acs.org.

## AUTHOR INFORMATION

### Corresponding Author

*E-mail: sunryu@postech.ac.kr

### Author Contributions

The manuscript was written through the contributions of all authors. All authors have given approval to the final version of the manuscript.

**Notes**

The authors declare no conflict of interest.

**ACKNOWLEDGMENTS**

This work was supported by the National Research Foundation of Korea (NRF-2015R1A2A1A15052078 and NRF-2016R1A2B3010390). The authors also thank Dogyeong Kim for assistance in the reflectance measurements.

**FIGURES & CAPTIONS**

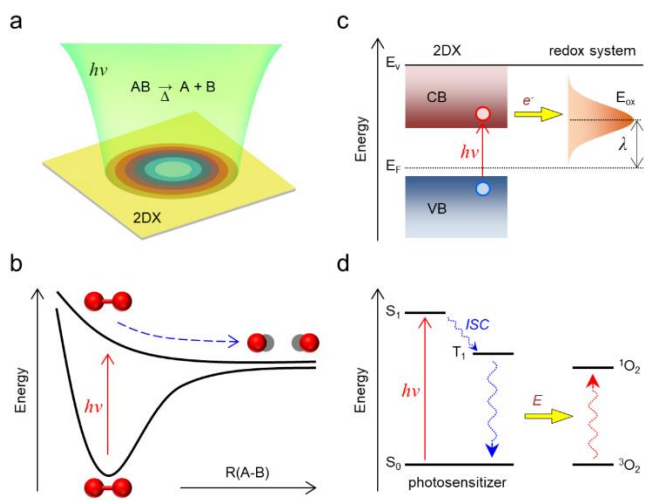

**Figure 1. Reaction mechanisms for various photoinduced events in 2DXs.** (a) Photothermal dissociation of a local bond (AB) in a 2DX is driven by the heat (Δ) that is converted from the absorbed photons (hv). (b) Direct photodissociation may occur when the chemical entity is photoexcited to a repulsive state with respect to its reaction coordinate (horizontal axis). (c) Hot electron-mediated mechanism is initiated by the electron photoexcited from a valence band (VB) to a conduction band (CB) but still below the vacuum level ($E_v$). The hot electron is transferred to an oxidized form of the redox couple at $E_{ox}$ that is located λ (reorganization energy) above the Fermi level ($E_F$). (d) In a sensitized photoreaction, the photon energy absorbed by 2DX is transferred to another nearby chemical entity ($O_2$) with or without a spin flip, which is followed by a chemical reaction driven by the excited $O_2$. Typically, the spin flip is preceded by intersystem crossing in a molecular photosensitizer as depicted in (d).

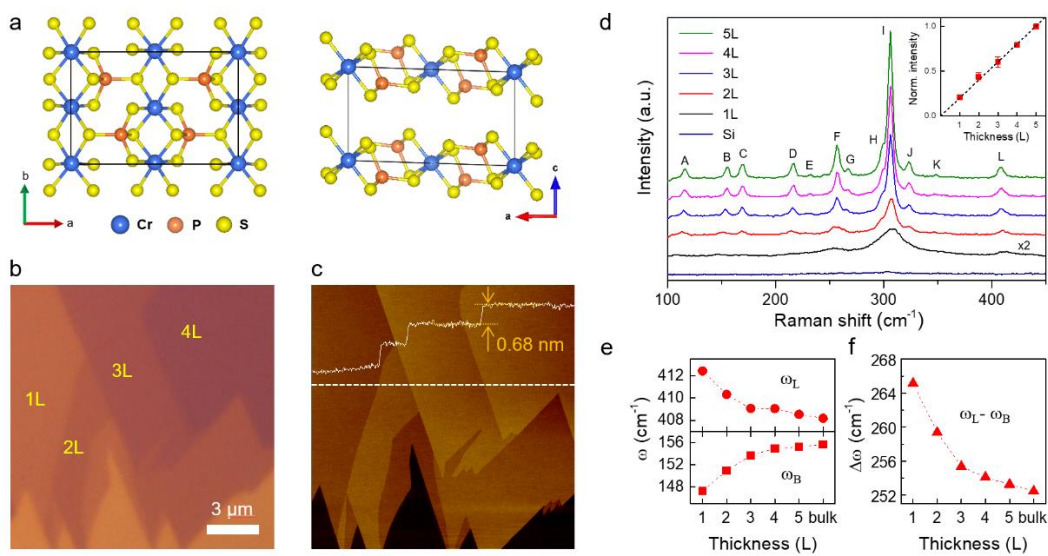

**Figure 2. Crystal structure, surface morphology, thickness-dependent photoreactivity and lattice vibrations of 2D CrPS₄.** (a) Ball-and-stick model of 2L CrPS₄ belonging to the space group $C_3^2$: top (left) and side views (right), where a unit cell is marked by black parallelograms and crystallographic axes are represented by three arrows. Blue, orange and yellow spheres designate Cr, P and S atoms, respectively. (b) Optical micrograph of 1 ~ 4L CrPS₄ on a SiO₂/Si substrate. (c) AFM topographic image obtained from the same area shown in (b). The height profile (solid line) along the white dashed line shows that the interlayer distance is ca. 0.68 nm. (d) Raman spectra of 1 ~ 5L CrPS₄ obtained in an Ar atmosphere. Unlike multilayers, Raman signal was not detected for unprotected 1L because of immediate photodegradation, which indicated a severe dependence of photoreactivity on thickness. The 1L spectrum was obtained from one sample encapsulated with 10 nm-thick Al₂O₃ layers. The inset presents the intensity of the F peak ($I_F$) as a function of thickness. (e) Frequencies of the B and L peaks respectively showing upshift and downshift with increasing thickness. (f) Frequency difference between the L and B peaks shown as a function of thickness.

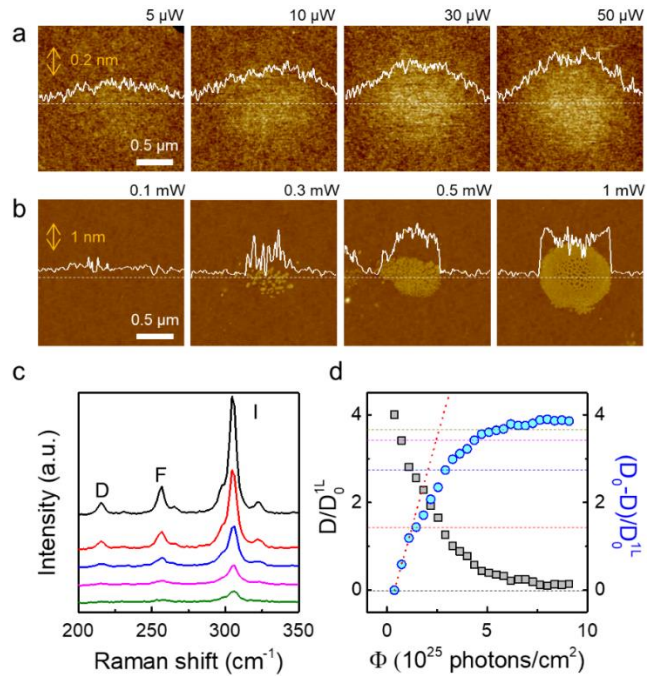

**Figure 3. Distinctively high photo-susceptibility of CrPS₄ and its quantification.** (a & b) Non-contact AFM topographic images of 4L CrPS₄ (a) and 4L MoS₂ (b) obtained after photoirradiation for 200 s at various average power in the ambient air. The height profiles (solid lines with vertical scale bars) obtained along the dashed lines show protrusions induced by the photoreactions for both materials. (c) Raman spectra of 4L CrPS₄ obtained with increasing exposure time ($t_{ex}$): (top to bottom) 20, 80, 160, 240 and 320 s at 0.2 mW in the ambient air. The spectra were offset for clarity. (d) Normalized effective thickness (black squares) of intact CrPS₄ ($D / D_0^{IL}$) determined from $I_F$ as a function of photon fluence ($\Phi$) for 4L CrPS₄ at 0.2 mW in the ambient air. In the right ordinate, the normalized extent of reaction ($\xi = (D_0 - D) / D_0^{IL}$) is given by blue circles with a tangent to the early stage for determination of reaction rates (see text for its details). Each of the horizontal dashed lines corresponds to the Raman spectrum of the same color in (c).

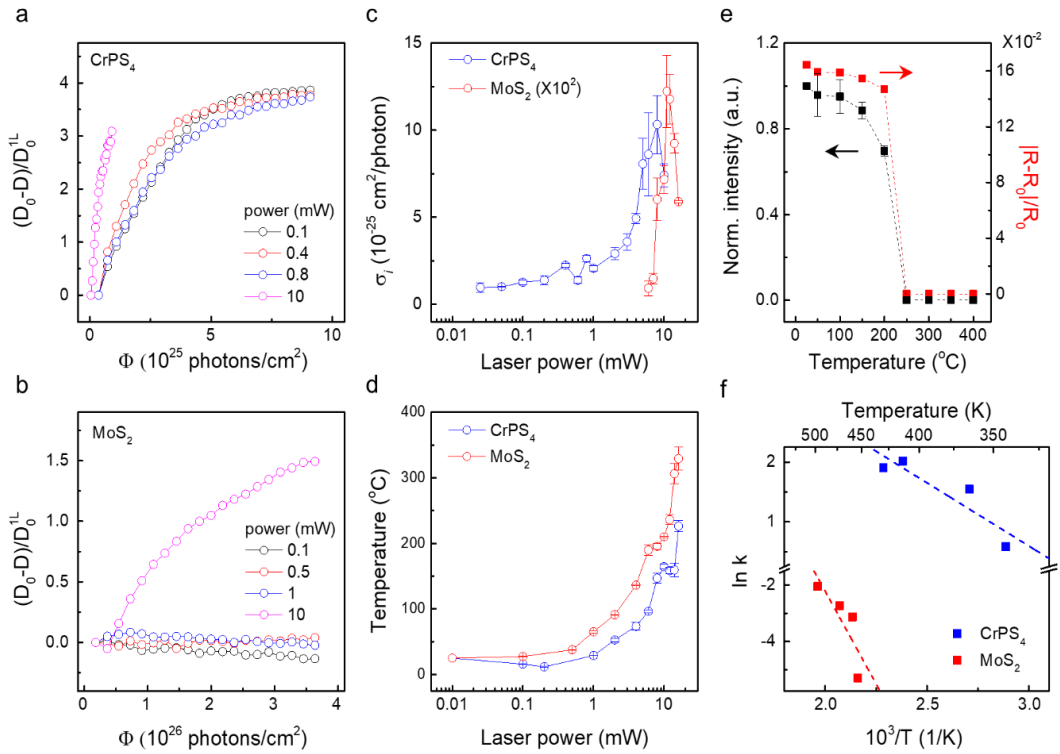

**Figure 4. Crossover between photochemical and photothermal oxidation mechanisms.** (a & b) Normalized extent of reaction ($\xi$) for 4L CrPS$_4$ (a) and 4L MoS$_2$ (b) obtained with various average power in the air. (c) Initial reaction cross section ($\sigma_i$) of both samples determined for a wide range of average power. (d) Temperature of samples ($T_{opt}$) determined in-situ during photoirradiation using Raman thermometry in a range of average power. (e) Thermal oxidation of 4L CrPS$_4$ probed with Raman intensity ($I_F$) and red-channel optical contrast ($|R - R_0| / R_0$) with increasing oxidation temperature in the air. (f) Arrhenius plots for photothermal reactions of 4L CrPS$_4$ and MoS$_2$. CrPS$_4$ showed ~6 times smaller activation energy than MoS$_2$.

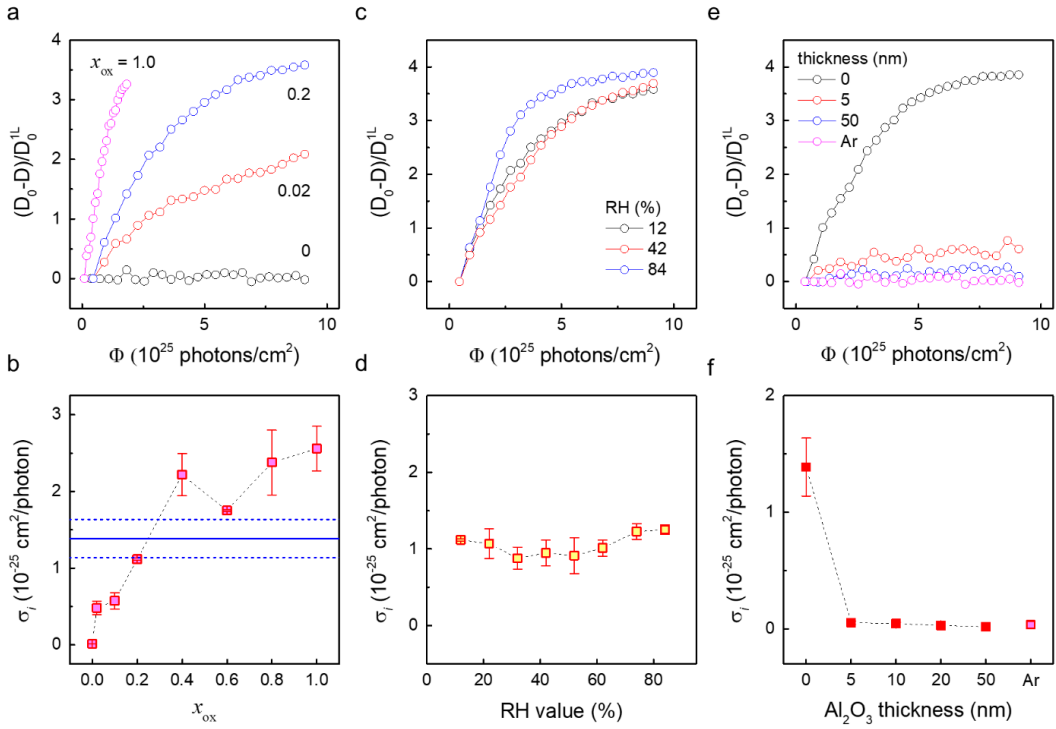

**Figure 5. Role of oxygen and water in photooxidation of CrPS₄.** (a & b) Dependence of the reaction rate on the mole fraction of $O_2$ in Ar gas ($\chi_{ox}$): $\xi$ of 4L CrPS₄ as a function of $\Phi$ obtained for various $\chi_{ox}$ (a) and $\sigma_i$ determined for a range of $\chi_{ox}$ (b). The solid blue line represents $\sigma_i$ obtained in the ambient air with the two blue dashed lines marking error bounds. (c & d) Dependence of the reaction rate on water vapor: $\xi$ of 4L CrPS₄ as a function of $\Phi$ obtained for various relative humidity (RH) (c) and $\sigma_i$ determined for a range of RH (d), where $\chi_{ox}$ was fixed at 0.2. (e & f) Effects of encapsulation on the reaction: $\xi$ of 4L CrPS₄ as a function of $\Phi$ obtained for samples with or without $Al_2O_3$ layers in the air (e) and $\sigma_i$ determined for different thicknesses of $Al_2O_3$ layers (f). For comparison, data obtained from unprotected 4L in an Ar gas is shown together.

**Graphic TOC**

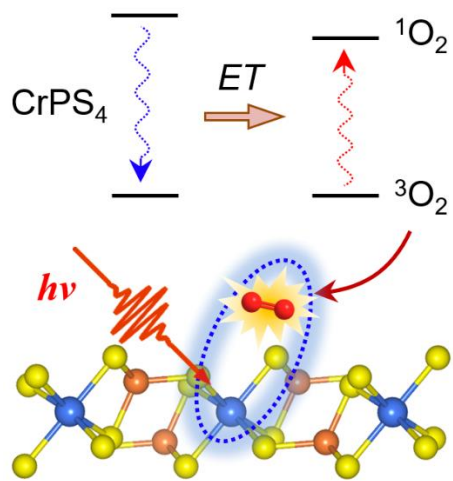